\documentstyle[preprint,aps]{revtex}
\begin{document}
\draft
%\tighten
\title{Influence of Dynamical Pauli Effect and \\ Dynamical 
Symmetry Breaking to Quantum Chaos}
\author{{\bf Ming-Tsung Lee$^a$, Wei-Min Zhang$^b$ and Cheng-Li Wu$^c$}}
\address{$^a$Department of Physics, National Taiwan University, Taipei 107, 
Taiwan, ROC \\
$^b$Institute of Physics, Academia Sinica, Taipei 115, Taiwan, ROC \\
$^c$Department of Physics, Chung-Yuan Christian University, Chung Li 
320, Taiwan, ROC}

\date{Oct. 22, 1997}

\maketitle

\begin{abstract}
In this paper, we study the influence of quantum effects to chaotic
dynamics, especially the influence of Pauli effect and dynamical symmetry
breaking to chaotic motions. We apply the semiquantal theory to the Sp(6)
fermion symmetry model in nuclear collective motion. We demonstrate that
quantum chaos appears when dynamical symmetry is broken. We further show
that dynamical Pauli effect enhances quantum chaos.
\end{abstract}
\vskip .2in
\centerline{PACS: 05.45.+b, 21.60.-n, 24.60.Lz, 03.65.Sq}
\vskip .2in

\section{\bf INTRODUCTION}

Quantum chaos is one of the most challenge research topics in both 
theoretical and experimental physics. In the past decade, in order 
to understand the concept of quantum chaos, people have been focusing 
on seeking the generic behavior in quantum spectra and wave functions 
of various dynamical systems whose classical dynamics are chaotic. 
In these investigations, two significant features have been found. 
One of them is the relation between the quantum energy level statistical 
distributions of the classical chaotic systems and the GOE/GUE 
spectral statistical universality classes of the random matrix 
theory \cite{Bohigas}. The other feature is the existence of ``scars" 
in individual eigenstate wave functions, i.e., the appearance 
of strong localization of wave functions along the unstable periodic 
orbits in chaotic systems \cite{Gutzwiller}. However, the dynamical 
mechanism of quantum manifestation of chaos is still not well understood 
so far.

In order to study the dynamical mechanism of quantum manifestation of 
chaos, one of us developed an approach, called the semiquantal approach 
\cite{zhang95}. The main motivation of this approach is to provide a 
framework to study how quantum fluctuations, quantum correlations and 
other quantum phenomena manifest themselves in the classical trajectories. 
This approach differs from the above mentioned investigations. It starts 
from quantum systems without reference to classical limit and it
directly explores the dynamics of quantum effects in classical chaotic
trajectories. Therefore, it provides a systematic way to explore 
the intrinsic mechanism of quantum manifestation of chaos\cite
{zhang90a,zhang94,Pat94,Hu97,Hu97a}.

The general theoretical framework of the semiquantal approach is 
the stationary phase approximation of the generalized coherent state 
path integral \cite{zhang90}. Each set of generalized coherent states 
associated with the dynamical group $G$ of a quantum system forms an 
over-complete set of states for an irreducible representation space 
of $G$. The coherent states are also in one-to-one corresponding to a 
geometrical space which has a symplectic structure. We called this 
geometrical space the quantum phase space. Based on path integral 
formulation, coherent states then embed each quantum system in a 
quantum phase space, regardless of whether the system has 
a classical counterpart or not. The stationary phase approximation 
further results in a set of dynamical equations of motion on this 
quantum phase space, which are similar to the classical equations of 
motion. But the Hamiltonian function governed these dynamical equations 
of motion is the expectation values of the Hamiltonian operator in the 
generalized coherent states, which contain the leading 
quantum fluctuations and correlations. We called these 
equations of motion the semiquantal dynamical equations. This formulation
allows us to directly explore the dynamical effects of quantum fluctuations
and correlations to chaos \cite{zhang95}.

Based on this formulation of semiquantal dynamics, we will study 
in this paper the dynamical effect of the Pauli exclusive principle to 
chaotic motions. We shall take the Sp(6) model, the Sp(6) symmetry of the
fermion dynamical symmetry model in nuclear many-body systems 
\cite{fdsm}, as our example. This model is a microscopic description of 
nuclear collective dynamics with fermions as the building blocks. A 
significant feature of this model is that the Pauli principle of fermions 
has a crucial influence on the low-lying nuclear collective modes, which 
manifests themselves in experimental data\cite{Wu88}. When we define a 
set of canonical deformation variables (the quantum phase space) 
associated with the Sp(6) coherent states, the connection between 
Bohr-Mottelson geometrical model \cite{Bohr} and the Sp(6) model can be 
built \cite{zhang88}. The Pauli effect in the Sp(6) model itself 
manifests in the nuclear deformation in terms of some constraints 
on the nuclear surface (deformation) 
variables \cite{zhang89}.  Within the semiquantal framework, the 
Poincare's picture of chaos can be applied to describe the possible 
chaotic dynamics in nuclear collective motion in the aforementioned 
quantum phase space. Therefore, dynamical Pauli effect to chaos can be 
explicitly studied. Our result shows that dynamical Pauli effect
enhances the pairing correlations (quantum correlation) and inhibits 
nuclear deformation. This gives rise to the breaking of dynamical
symmetry and therefore leads to the occurrence of chaos.

The paper is organized as follows: In Sec.~2, the nuclear shape in terms 
of the microscopic fermion degrees of freedom is parameterized through 
the Sp(6) coherent states. The coherent state of the Sp(6) model in the
nuclear body frame is constructed so that the quantum phase space of the 
Sp(6) model is connected with the nuclear geometrical model. In Sec.~3, 
the semiquantal dynamical equations of motion for nuclear deformation is
derived. With the consideration of time-reversal invariance and valence
nucleon number conservation, the resulting equations of motion are reduced
to a set of Hamilton-like equations in a four-dimensional quantum phase 
space, which describes the nuclear surface motions. In Sec.~4, we 
explore the geometrical manifestation of dynamical Pauli effect and
numerically examine the dynamical effect of the Pauli principle and 
dynamical symmetry breaking in semiquantal framework, from which we 
find that quantum chaos appears when dynamical symmetry is broken and  
dynamical Pauli effect enhances quantum chaos.  Finally, a summary is 
presented in Sec.~5.

\section{Phase space parameterization of nuclear shape and geometrical 
interpretation}

We begin with the fermion dynamical symmetry Sp(6) model \cite{fdsm}
which describe the nuclear collective motions. The Sp(6) group is 
spanned by the collective pairing operators $\{\hat S, \hat D_\mu\}$ and 
the collective particle-hole quadrupole operators $\{ \hat P_{\lambda 
\mu}; \lambda=0,1,2; \mu=-\lambda, ..., \lambda \}$. The model
Hamiltonian operator is given by
\begin{equation}
	\hat{H}=\epsilon \hat{n}+G_0\hat{S}^{+}\hat{S}+ \sum_{l=1,2} b_l 
		\hat{P}_l\cdot \hat{P}_l \, ,  \label{eq32} 
\end{equation}
where $\hat{n}$ is the particle number operator. The coefficient $\epsilon,
G_0$ and $b_l$ represent the strength of the internal energy, effective
pairing and multiple interactions, respectively. It is worth noticing that
the $D$-pairing interactions are not included in Eq.~(\ref{eq32}) because 
such interactions can be replaced by the $S$-pairing and multiple 
interactions in the Sp(6) symmetry\cite{fdsm}. Also, in order to study 
chaotic dynamics of nuclear deformation under the Sp(6) model, we should 
only consider the dynamics of different intrinsic bands rather then the 
collective motion within a single band. Therefore, we can let $b_1=b_2$ 
in Eq.~(\ref{eq32}). 

To study the nuclear surface motions in the Sp(6) model within the 
semiquantal framework, one must first construct the corresponding 
quantum phase space. 
It has been shown that such a quantum phase space is uniquely realized 
by the coset space USp(6)/U(3) through the Sp(6) coherent states. (The 
detailed construction of the coherent states of the Sp(6) model can be 
found in Ref.\cite{zhang88}. It has also been roughly summarized in 
Appendix A.)

In the following subsections, we will parametrize the nuclear shape in the 
above mentioned quantum phase space by visualizing nuclear deformation 
to be a geometrical description of the microscopic Sp(6) fermion 
model. As it is well-known, the dynamics of nuclear deformation has 
time reversal invariance. Meantime, it is also convenient to describe
nuclear deformation in the nuclear intrinsic frame in which adequate 
deformed variables can be chosen as the coordinates of the Sp(6) quantum 
phase space. 

\subsection{The time reversal invariant deformed coherent states}

First, we shall look at the connection of the Sp(6) Coherent States (CS)
with nuclear deformation. In the macroscopic picture of the nuclear
geometrical model, the rotational and vibrational energy spectrum of nuclei
are described in according to the nuclear surface motions. The nuclear
surface can be defined by
\begin{equation}  \label{eq1}
	 R(t,\theta ,\phi )=R_0 \Bigg( \sum_{\lambda =0,1,2, \\\mu =-\lambda,
		\cdots ,\lambda} \alpha_{\lambda \mu}^{*}(t) 
		Y_{\lambda \mu } (\theta ,\phi) \Bigg) +R_c \, ,
\end{equation}
where $R_0$ is the maximum thickness of a nucleus, $R_c$ 
the radius of frozen core, and $Y_{\lambda\mu }( \theta ,\phi)$ 
the spherical harmonic function. In Eq.~(\ref{eq1}), the 
quadrupole terms (with $\lambda=2$) play the dominant role in the 
description of nuclear deformation. They determine the strength of 
nuclear surface motions. The monopole term ($\lambda =0$) describes 
the shell thickness of the undeformed compressible nucleus. When the
center of nuclei is fixed in a reference frame (no translation), the 
effect of $\lambda =1$ term can be ignored. Besides, all the surface 
variables in Eq.~(\ref{eq1}) satisfy the time reversal symmetry so that 
\begin{equation}  \label{eq2}
	\alpha _{\lambda \mu }^{*}(t)=(-1)^\mu \alpha _{\lambda -\mu }(t)
		\, .
\end{equation}

To generalize this concept into the framework of a microscopic 
description, we introduce a deformation matrix ${\bf \rho }$, 
\begin{equation}
	{\bf \rho }=-\left( \alpha _{00} ( t) {\bf I}_{3\times 3}
	{\bf +} 2\sum_{\mu =-2}^2\alpha _{2\mu }^{*}(t){\bf P}_{2\mu }\right)  
	\, , \label{eq3} 
\end{equation}
which is related to the quadrupole momentum in the Sp(6) model,
where ${\bf I}_{3\times 3}$ is a 3$\times $3 identity matrix; and 
${\bf P}_{2\mu }$ the submatrix block of the quadrupole momentum 
operator $\hat{P}_{2\mu }$ in the $6 \times 6$ matrix representation
of the Sp(6) group. The detailed expressions can be found in Appendix 
A of Ref.\cite{zhang88}. These submatrices ${\bf P}_{\lambda \mu }$
form an irreducible representation of the unitary subgroup U(3). 
They determine the deformation properties of nucleus. The 
surface variables $\alpha_{\lambda \mu }$ still obey the time reversal 
invariance condition (\ref{eq2}), and the coefficient $\alpha_{00}(t)$ 
is chosen to be negative. This is because the shell thickness of the
undeformed nucleus is positive, and also the quadrupole momentum must 
also be positive when the particle number is less than $\Omega $, 
where $\Omega $ is the pair degeneracy of normal-parity levels in 
a major shell (i.e. $\Omega _1$ in the Sp(6) model). In this paper 
we only consider the case of valence particle number $n\leq \Omega 
$. For the case $n>\Omega $, an analogous discussion can be made 
because of the particle-hole symmetry.

Next, we shall build the connection between the surface variables 
$\alpha_{\lambda \mu }(t)$ and the parameters in the Sp(6) coherent 
states. As one can find in Ref.\cite{zhang88} (also see Appendix A),
the Sp(6) coherent state $| \eta \rangle$ is completely determined 
by a $3\times 3$ parameter matrix {\bf X}. Meantime, the 
$3\times 3$ hermitian matrix $\sqrt{{\bf XX^{+}}}$ can be spanned by 
${\bf P}_{2\mu }$: 
\begin{equation}
	\sqrt{{\bf XX^{+}}}=\sum_{\lambda =0,2,\\\mu =-\lambda ,..,
		\lambda }\beta_{\lambda \mu }^{*}(\chi _{_{\lambda 
		^{^{\prime }}\mu ^{^{\prime }}}}(t) )
		{\bf P}_{\lambda \mu } \, ,  \label{eq4}
\end{equation}
where ${\bf P}_{\lambda \mu }$ obey the condition: ${\bf P}_{\lambda 
-\mu}=(-1)^\mu {\bf P}^t_{\lambda \mu }$.  Furthermore, the 
hermitian property of the matrix $\sqrt{{\bf XX^{+}}}$ implies that
\begin{equation}
	\qquad \beta _{\lambda -\mu}^{*}=(-1)^\mu \beta _{\lambda \mu }  
\end{equation}
which corresponds to the time reversal invariance.

Without any loss of generality, we can set the matrix $\sqrt{{\bf 
XX^{+}}}$ to be equal to the deformation matrix ${\bf \rho }$ by letting 
$\beta _{\lambda \mu }(t)=(-1)^\mu \alpha _{\lambda \mu }(t)$. The 
connection between the surface variables $\alpha _{\lambda \mu }(t)$ and 
the coherent state parameters in $| \eta \rangle $ can then be 
carried out. The properties of nuclear deformation determined by the 
Sp(6) CS are described by Eqs.~(\ref{eq2}-\ref{eq3}) and the 
requirement $\sqrt{{\bf XX}^{+}} \equiv {\bf \rho }$. The explicit form 
of the matrix $\sqrt{{\bf XX}^{+}}$ is given as
\begin{equation}
	\sqrt{{\bf XX}^{+}}\equiv {\bf \rho } \\ =-\left[ 
	  \begin{array}{ccc}
		\alpha _{00}+\sqrt{\frac 12}\alpha _{20} & \sqrt{\frac 32}
		\alpha _{21}^{*} & \sqrt{3}\alpha _{22}^{*} \\ 
		\sqrt{\frac 32}\alpha _{21} & \alpha _{00}-\sqrt{2}
		\alpha _{20} & -\sqrt{\frac 32}\alpha _{21}^{*} \\ 
		\sqrt{3}\alpha _{22} & -\sqrt{\frac 32}\alpha _{21} & 
		\alpha _{00}+\sqrt{\frac 12}\alpha _{20}  \end{array}
		\right] \,.  \label{eq5}
\end{equation}

\subsection{The deformed coherent states in the intrinsic frame (ICS)}

In order to make the dynamics of nuclear deformation manifestation, 
it is convenient to express the nuclear deformation parameters in 
the nuclear intrinsic frame (i.e. the nuclear body frame). We call 
the CS describing the nuclear deformation in the nuclear intrinsic 
frame the intrinsic coherent states (ICS). In the intrinsic frame, 
the deformation matrix is positive and diagonal. To deduce the ICS, a 
basic idea is to decompose the previous spherical tensor operators into 
proper cartesian tensor operators.

The cartesian tensor operator $\hat{a}_{ij}$ are defined as follows: 
\begin{equation}
	\hat{a}_{ij}=\sum_{lm=1,2,3}\left( U_{il}^{}\hat{B}_{lm}
	U_{mj}^{+}\right) ~~, ~~~ i,j=1,2,3 \, ,
\end{equation}
where the index $ 1\rightarrow x,2 \rightarrow z,3 \rightarrow y$ 
for $\hat{a}_{ij}$. Such an index assignment corresponds to a choice 
of the left-hand coordinate system in order to be consistent with the 
notations used in Ref.\cite{zhang88}.
The operators $\hat{B}_{lm}$ are uncoupled particle-hole operators 
in the Sp(6) model. The connection between $\hat{B}_{lm}$ and 
$\hat{P}_{\lambda \mu }$ can be found in Ref.\cite{zhang88}. The 
transformation matrix $U$ is defined by: 
\begin{equation}
	U=\frac 1{\sqrt{2}}\left[ \begin{array}{ccc}
		1 & 0 & -1 \\ 0 & \sqrt{2} & 0 \\ -i & 0 & -i \end{array}
		\right] \,.
\end{equation}
With the above cartesian tensor operators, the rank-one spherical 
tensor operators can be expressed in terms of $\hat{a}_{ij}$ ,
\begin{eqnarray}
	\hat{P}_{11} &=& \frac{\sqrt{3}}4\left( \hat{a}_{zx}-\hat{a}_{xz}
		+i~\hat{a}_{zy}-i~\hat{a}_{yz}\right) \nonumber \, ,\\ 
	\hat{P}_{1-1}&=& \frac{\sqrt{3}}4\left( \hat{a}_{zx}-\hat{a}_{xz}
		+i~\hat{a}_{yz}-i~\hat{a}_{zy}\right) \nonumber \, , \\ 
	\hat{P}_{10}&=& i\sqrt{\frac 38}\left( \hat{a}_{yx}-\hat{a}_{xy}\right)
		\,,  \label{eq8}
\end{eqnarray}
and the rank-two spherical tensor operators $\hat{P}_{2\mu }$ (quadrupole
momentum operators) ,
\begin{eqnarray}
	\hat{P}_{20} &=& \sqrt{\frac 18}\left( \hat{a}_{xx}-2\hat{a}_{zz}
		+\hat{a}_{yy} \right)  \nonumber \, , \\ 
	\hat{P}_{21} &=& -\frac{\sqrt{3}}4\left( \hat{a}_{zx}+\hat{a}_{xz}
		+i~\hat{a}_{yz}+i~\hat{a}_{zy}\right) \nonumber \, , \\ 
	\hat{P}_{2-1}&=& \frac{\sqrt{3}}4\left( \hat{a}_{zx}+\hat{a}_{xz}
		-i~\hat{a}_{yz}-i~\hat{a}_{zy}\right) \nonumber \, , \\ 
	\hat{P}_{22} &=& \frac{\sqrt{3}}4\left( -\hat{a}_{xx}+\hat{a}_{yy}
		-i~\hat{a}_{xy}-i~\hat{a}_{yx}\right) \nonumber \, , \\ 
	\hat{P}_{2-2}&=& \frac{\sqrt{3}}4\left( -\hat{a}_{xx}+\hat{a}_{yy}
		+i~\hat{a}_{xy}+i~\hat{a}_{yx}\right) \, . \label{eq9}
\end{eqnarray}

Let us now consider the matrix representation in terms of 
$\hat{a}_{ij}$ as the basis, 
\begin{equation}
	\hat{a}_{ij}\longrightarrow Rep(\hat{a}_{ij})=e_{ij}=\left[ 
		\begin{array}{cc} E_{ij} & 0 \\ 0 & -E_{ji}
			\end{array} \right] \,,
\end{equation}
where $E_{ij}$ is a $3\times 3$ matrix with +1 in the $i$-th row and 
the $j$-th column, and zero everywhere else. The deformation matrix 
${\bf \rho }$ in this cartesian basis is then expressed as
\begin{equation}
	{\bf \rho }=-\left[ \begin{array}{ccc} 
	   \begin{array}{l} \alpha _{00}+\sqrt{\frac 12}\alpha _{20} \\ 
	    ~~~~ -\sqrt{\frac 34}(\alpha _{22}+\alpha _{22}^{*}) \end{array}
		& \sqrt{\frac 34}(\alpha _{21}+\alpha _{21}^{*}) & 
		i\sqrt{\frac 34}(\alpha_{22}-\alpha _{22}^{*}) \\ 
		\sqrt{\frac 34}(\alpha _{21}+\alpha _{21}^{*}) & \alpha _{00}
		-\sqrt{2}\alpha_{20} & -i\sqrt{\frac 34}(\alpha_{21}
		-\alpha _{21}^{*}) \\ i\sqrt{\frac 34}(\alpha_{22}
		-\alpha _{22}^{*}) & i\sqrt{\frac 34}(\alpha_{21}^{*}
		-\alpha _{21}) & 
	  \begin{array}{l} \alpha _{00}+\sqrt{\frac 12}\alpha _{20} \\ 
	   ~~~~ +\sqrt{\frac 34}(\alpha _{22}+\alpha _{22}^{*})\end{array}
		\end{array} \right] \, , \label{eq17}
\end{equation}
where the parameters $\{ \alpha _{\lambda \mu }\} $ maintain 
the tensorial structure under rotational transformation, i.e.
\begin{equation}
	R^{-1}( \omega ) \ {\bf \rho }\left( \alpha _{\lambda 
		\mu}\right) R\left( \omega \right) ={\bf \rho }\left( 
		\alpha _{\lambda \nu}^{^{\prime }}\right) ,
\end{equation}
and $\alpha'_{\lambda \nu }$ is given by
\begin{eqnarray}
	\alpha'_{\lambda \nu } &=&\sum_{\mu =-\lambda}^\lambda 
	\alpha_{\lambda \mu } D_{\mu \nu }^\lambda ( \omega )  \, , \\ 
	\alpha_{\lambda \nu }^{'*} &=&(-1)^\nu \alpha'_{\lambda -\nu} \, ,
\end{eqnarray}
with $R ( \omega ) $ being a Euler rotational matrix, $\omega $ 
the Euler angular, and $D_{\mu \nu }^\lambda ( \omega ) $ the 
Wigner-$D$ function. One can always find an Euler rotation
$R( \omega ) $ such that the nucleus is described in an 
intrinsic frame in which ${\bf \rho }$ is diagonal (i.e. $\alpha'_{21} 
=0$ and $\alpha' _{22}$ is real): 
\begin{equation}	\label{eq15}
	{\bf \rho }_{in}=\left[ \begin{array}{ccc}
		\left| S_x\right| & 0 & 0 \\ 0 & \left| S_z\right| & 0 \\ 
		0 & 0 & \left| S_y\right|  \end{array} \right] \, .	
\end{equation}
Since ${\bf \rho }_{in}=\sqrt{{\bf X}_{in}{\bf X}_{in}^{+}}$, the matrix 
${\bf X}_{in}$ (the Sp(6) CS parameter matrix in the intrinsic frame) 
can be chosen as: 
\begin{equation}
	{\bf X}_{in}=\left[ \begin{array}{ccc} S_x & 0 & 0 \\ 
		0 & S_z & 0 \\ 0 & 0 & -S_y \end{array}
		\right] \, .  \label{eq24}
\end{equation}
The minus sign of $S_y$ in the above equation is due to the use of
the right hand coordinate system.  The explicit form of the parameters 
$S_i$ is given as follows, 
\begin{eqnarray}
	S_x &=& -\left( \alpha _{00}+\frac 1{\sqrt{2}}\alpha _{20}
		-\sqrt{3}\alpha _{22}\right) e^{-if_x (\alpha _{00},
		\alpha_{20},\alpha _{22};\pi _{00},\pi _{20},\pi _{22})} 
		\nonumber \, , \\ 
	S_z &=& \left( -\alpha _{00}+\sqrt{2}\alpha _{20}\right) e^{-if_z
		( \alpha_{00},\alpha _{20},\alpha _{22};\pi _{00},\pi _{20},
		\pi _{22}) } \nonumber \, , \\ 
	S_y &=& -\left( \alpha _{00}+\frac 1{\sqrt{2}}\alpha _{20}
		+\sqrt{3}\alpha_{22}\right) e^{-if_y( \alpha _{00},
		\alpha _{20},\alpha _{22};\pi_{00},\pi _{20},\pi _{22}) }
		\, ,  \label{eq23}
\end{eqnarray}
where the detailed functions of $f_x$, $f_z$ and $f_y$ can be found in 
Appendix C. In the intrinsic frame, the deformation degrees of freedom 
($\sqrt{{\bf 
XX}^{+}}$) are reduced to three because the intrinsic frame is chosen to 
be along the three principal axes of the nucleus. Together with the 
corresponding 
conjugate momentum variables, the phase space of nuclear deformation
is a six-dimensional space. The six independent variables are now
completely determined by the three complex parameters of ${\bf X}_{in}$ 
in ICS.

Based on ICS, the nuclear deformation is described by the variables 
$|S_i| $ only. The physical picture of the variables $| S_i|$
can be seen from Eq.~(\ref{eq1}) which can be reexpressed in 
terms of cartesian coordinates as follows: 
\begin{eqnarray}
	R(t,x,z,y) &=& R_c + R_0 \Big(| \alpha _{00}| 
		+\alpha _{xx}\frac{x^2}{r^2}+\alpha _{yy}\frac{y^2}
		{r^2}+\alpha _{zz}\frac{z^2}{r^2} \nonumber \\ 
	& & ~~~~~~~~~~~  +2\alpha _{xz}\frac{xz}{r^2}+2\alpha_{xy}
		\frac{xy}{r^2}+2\alpha _{yz}\frac{yz}{r^2} \Big) \, ,
\end{eqnarray}
where $r=\sqrt{x^2+z^2+y^2}$. Since the intrinsic frame is chosen  
to be the three principal axes of nucleus in the intrinsic frame, 
the terms $\alpha _{xz}$, $\alpha_{xy}$ and $\alpha _{yz}$  vanish. The 
deformation matrix then becomes: 
\begin{equation}
	{\bf \rho }_{in}=-\alpha _{00}(t){\bf I}_{3\times 3}
		+\sum_{i=x,y,z} \alpha_{ii}(t) {\bf A}_{ii} \, , \label{eq26}
\end{equation}
where ${\bf A}_{ii}$ is the subblock matrix in the matrix representation of 
$\hat{a}_{ij}$, and  
\begin{equation}
	\alpha _{ii}(t)=\alpha _{00}(t)+ | S_i(t)| \, .
\end{equation}
According to Eq.~(\ref{eq26}), the length of the nucleus along the 
$i$th-axis is given by $R_0 | S_i(t) | +R_c$. In other words, 
$\{ S_i \} $ are the variables measuring nuclear deformation. These 
are exactly the dynamical variables we want to use to study the nuclear 
surface motions.

Once we have constructed the Sp(6) CS in the intrinsic frame, we can
now derive the semiquantal dynamical equations of the nuclear collective
motions.

\section{\bf THE SEMIQUANTAL DYNAMICS FOR Sp(6) MODEL}

The semiquantal dynamics of the Sp(6) model is determined by the
time-dependent variational approach to the effective action (i.e. the
stationary phase approximation of the generalized CS path integral): 
\begin{equation}
	S=\int dt \langle \text{ICS} | i\partial _t-\hat{H} 
		| \text{ICS} \rangle \, .
\end{equation}
The result of such a variation is a set of Hamilton-like equations 
\begin{eqnarray}
	i\hbar \frac \Omega 3\dot{S}_i^{*} &=& -\frac{\partial \langle 
		\hat{H} \rangle }{\partial S_i} \nonumber \, , \\ 
	i\hbar \frac \Omega 3\dot{S}_i &=& \frac{\partial \langle \hat{H}
		\rangle }{\partial S_i^{*}} \qquad ,\qquad i=x,y,z \, .  
		\label{eq31}
\end{eqnarray}
These equations of motion are the semiquantal dynamical equations 
which approximately describe the quantum dynamics of nuclear deformation
in the Sp(6) model. The word ``approximately'' (or semiquantal) means 
that this is an 
approximation of the Sch\" odinger equation described by the
Hamilton-like equations of motion but it is derived purely from quantum
mechanics. And the Hamiltonian function $\langle \hat{H} \rangle = 
\langle \text{ICS} | \hat{H} | \text{ICS} \rangle $ 
contains all the leading order quantum correlations. In nuclear 
many-body systems, this set of equations of motion is indeed 
the familiar time-dependent Hartree-Fock-Bogoliubov equations under 
the constraint of dynamical symmetry group Sp(6) \cite{zhang89}.

Note that Eq.~(\ref{eq31}) contains six equations of motion. However, 
not all these six equations of motions are independent. This is because 
for a given nucleus, the valence nucleon number is fixed: 
\begin{equation}  \label{fvnm}
	n \equiv \langle \hat{n} \rangle =\frac{2\Omega }3 \left(|S_x|^2
		+| S_y| ^2+ | S_z|^2\right) = 2\Omega \left(
		\alpha _{00}^2+\alpha _{20}^2+2\alpha _{22}^2\right) \, .
\end{equation}
In order to solve Eq.~(\ref{eq31}), the constraint of the fixing valence
nucleon number must be imposed. In the next subsection, we shall 
first calculate the Hamiltonian function, and then introduce several 
canonical transformations to eliminate this constraint.

\subsection{The Hamiltonian function of the Sp(6) model}

The expectation value of the Hamiltonian operator (\ref{eq32}) in the
ICS is defined by
\begin{equation}  \label{eq33}
	\langle \hat{H} \rangle =\epsilon \langle \hat{n} \rangle 
	+G_0 \langle \hat{S}^{+}\hat{S} \rangle +b_2\sum_{l=1,2}
	\langle \hat{P}_l\cdot \hat{P}_l \rangle \, .
\end{equation}
The detailed form will be given later. Since the valence nucleon
number is conservative, the first term in Eq.~(\ref{eq33}) can be removed.
Hence, Eq.~(\ref{eq33}) can be simplified as 
\begin{equation}
	\langle \hat{H} \rangle =E_{[\text{SU(2)}]}+E_{[\text{SU(3)}] } \, ,
\end{equation}
with
\begin{eqnarray}
	E_{[\text{SU(2)}] }&=& G_0 \langle \hat{S}^{+}\hat{S} \rangle 
		\nonumber \, , \\ 
	E_{[ \text{SU(3})] }&=&b_2\sum_{l=1,2} \langle \hat{P}_l\cdot 
		\hat{P}_l\rangle \, ,	\label{cseve}
\end{eqnarray}
where the coherent state expectation values can be explicitly calculated 
in ICS, 
\begin{eqnarray}
	\langle \hat{S}^{+}\hat{S} \rangle &=& { \Omega \over 
		3} \left( {n \over 2} + \Big( 1 - {\Omega \over 3} \Big)
		K_2 + {2 \Omega \over 3} K_1 \right)  \nonumber \, , \\
	\langle \hat{P}_1\cdot \hat{P}_1 \rangle &=&\frac{3n}4-
		\frac \Omega 2K_3-\frac \Omega 2K_1 \nonumber \, , \\ 
	\langle \hat{P}_2\cdot \hat{P}_2 \rangle &=& \frac{7n}4-
		\frac{3n^2}{4\Omega }+{\Omega \over 2}K_1+ {\Omega \over
		3}K_2\Bigg({2\over 3} -1\Bigg) + {\Omega \over 3}
		K_3 \Bigg({1\over 2} - {2\Omega \over 3}\Bigg)
		\, ,	\label{csev1}
\end{eqnarray}
and 
\begin{eqnarray}
	K_1&=&\frac 12\sum_{i\neq j}\left[ \sqrt{1-| S_i| ^2}
		\sqrt{1-| S_j| ^2}S_iS_j^{*}\right] \nonumber \, , \\
	K_2&=& \sum_i | S_i| ^4  \nonumber \, , \\ 
	K_3&=& {1\over 2}\sum_{i\neq j}\left[| S_i| ^2 |S_j|^2\right]  \, .
\end{eqnarray}
In Appendix B, we have shown that the Sp(6) model in this parameter
representation (quantum phase space) contains all the general properties of
the nuclear geometrical model \cite{zhang88}.

\subsection{The semiquantal dynamical equations without constraint in the
Sp(6) model}

To remove the constraint of the fixed valence nucleon number, several
canonical transformations have to be made. Let the complex variables 
$S_i=q_i+ip_i$, where $q_i$ and $p_i$ are real variables. The Hamilton-like
equation (\ref{eq31}) can then be reexpressed as:
\begin{eqnarray}
	\frac{2\Omega }3\dot{p}_i &=& -\frac{\partial \langle \hat{H}\rangle}
		{\partial q_i} \nonumber \, , \\ 
	\frac{2\Omega }3\dot{q}_i &=& \frac{\partial \langle \hat{H}\rangle } 
		{\partial p_i} \, , \qquad i=x,y,z\,,
\end{eqnarray}
where the Planck constant $\hbar $ has been absorbed into the Hamiltonian
function. Next, we introduce a canonical transformation to generate the new 
canonical variables  $\{ {\cal P}_i,{\cal Q}_i \} $, and choose 
the second type generating functions as 
\begin{eqnarray}
	F_1^{(2)}&=&\sum_{i=x,y,z}\left( f_i(+)\text{ or }f_i(-)\right) 
		\nonumber \, , \\ 
	f_i(+) &=& \frac{q_i}2\sqrt{{\cal P}_i^2-q_i^2}+{{\cal P}_i^2 \over 2}
	   \left(\sin {}^{-1}\frac{q_i}{{\cal P}_i} - {\pi \over 2} \right)
		\nonumber \, , \\ 
	f_i(-) &=& -\frac{q_i}2\sqrt{{\cal P}_i^2-q_i^2}+{{\cal P}_i^2 \over 2} 
		\left( \cos{}^{-1}\frac{q_i}{{\cal P}_i} \right) \, .
\end{eqnarray}
This type of generating functions has eight assembles corresponding to 
positive or negative ${\it p}_i$, $i=x,y,z$. To be explicit, let us 
consider the special case, $F_1^{(2)}=\sum_i$ $f_i(+)$. 
From this canonical transformation, we can find the connections between 
the new and odd canonical variables,
\begin{eqnarray}
	p_i &=& \frac{\partial F_1^{(2)}}{\partial q_i}=\sqrt{{\cal 
		P}_i^2-q_i^2} \nonumber \, , \\
	{\cal Q}_i &=& \frac{\partial F_1^{(2)}}{\partial {\cal P}_i}=
		{\cal P}_i \left( \sin{}^{-1} {q_i \over {\cal P}_i}
		- {\pi \over 2} \right)  \, ,
\end{eqnarray}
or
\begin{eqnarray}
	q_i &=& {\cal P}_i\cos \frac{{\cal Q}_i}{{\cal P}_i} \nonumber \, , \\ 
	p_i &\equiv& | {\cal P}_i\sin \frac{{\cal Q}_i}{{\cal P}_i}| \, .
\end{eqnarray}
This transformation is valid only for positive $p_i$.  In another 
case $F_1^{(2)}=\sum_i$ $f_i(-)$, one can get the same transformation 
but for negative $p_i$. This is because in this case $p_i=-\sqrt{{\cal 
P}_i^2-q_i^2}=- | {\cal P}_i\sin \frac{{\cal Q}_i}{
{\cal P}_i}| $ and $q_i={\cal P}_i\cos \frac{{\cal Q}_i}{{\cal P}_i}$.
All the eight generating functions transform the old variables ($p_x,\
p_y,\ p_z$)=(+,+,+) or (+,+,--) or (--,--,+) ... into new variables (${\cal 
P}_x,\ {\cal P}_y,\ {\cal P}_z$), where (+,+,+) means that $p_x,\ p_y$ and $\
p_z$ are all positive. Their connections to the variables $S_i$ are given by 
\begin{equation}
	q_i^2+p_i^2=| S_i| ^2={\cal P}_i^2 \, ,
\end{equation}
and 
\begin{equation}
	S_i=q_i+ip_i={\cal P}_i\exp \left( -i\frac{{\cal Q}_i}{{\cal P}_i}
		\right) \, .  \label{eq42}
\end{equation}
Note that Eq.~(\ref{eq42}) implies ${\cal P}_i\neq 0$ (if ${\cal Q}_i \neq 
0$). In other words, the trajectories $( {\cal P}_x,{\cal P}_y,{\cal P}_z) $ 
cannot go through the
section ${\cal P}_i=0$. Therefore we have eight types of independent initial
conditions, namely, $( {\cal P}_x,{\cal P}_y,{\cal P}_z ) \in $ 
$\{ ( \pm ,\pm ,\pm) \} $. Next, we will
make the second canonical transformation. It transforms $( {\cal 
P}_i,{\cal Q}_i) $ into $( P_i,Q_i) $ such that $Q_z$ 
becomes an ignorable coordinate. We choose the second type generating 
function as

\begin{equation}
	F_2^{(2)}={\cal Q}_xP_x+{\cal Q}_yP_y\pm {\cal Q}_z\sqrt{P_z^2
		-P_x^2-P_y^2} \, .
\end{equation}
The connections between new and old variables are 
\begin{eqnarray}
	{\cal P}_i &=& \frac{\partial F_2^{(2)} (\pm) }{\partial {\cal 
		Q}_i} =P_i\qquad ,\qquad i=x,y  \nonumber \, , \\ 
	\frac{{\it Q}_z}{{\it P}_z} &=& \frac{{\cal Q}_z}{{\cal P}_z}
		\qquad ,\qquad \frac{Q_i}{P_i}=\frac{{\cal Q}_i}
		{{\cal P}_i}-\frac{{\cal Q}_z}{{\cal P}_z} \qquad ,\qquad 
		i=x,y \nonumber \, , \\ 
	{\cal P}_z^2 &=& P_z^2-P_x^2-P_y^2\qquad \text{or}\qquad P_z^2
		=\sum_i | S_i| ^2=\frac{3n}{2\Omega } \, .
\end{eqnarray}
By using these two steps of transformations, the relations between two sets
of canonical variables $\{ S_i,S_i^{*}\} $ and $\{P_i,Q_i\} $ are given by
\begin{eqnarray}
	| S_i|^2 &=& P_i^2\qquad ,\qquad i=x,y \nonumber \, , \\ 
	| S_z|^2 &=& \frac{3n}{2\Omega }-P_x^2-P_y^2 \nonumber 
		\, , \\ 
	S_xS_y^{*} &=& P_xP_y\exp \left[ -i\left( \frac{Q_x}{P_x}-\frac{Q_y}
		{P_y}\right) \right] \nonumber \, , \\ 
	S_iS_z^{*} &=& P_i\sqrt{\frac{3n}{2\Omega }-P_x^2-P_y^2}\exp \left( 
		-i\frac{Q_i}{ P_i}\right) ,i=x,y\text{ } \, .
\end{eqnarray}
The Hamiltonian function $H$ can be expressed now in terms of $(P_x, P_y,
Q_x, Q_y)$: 
\begin{eqnarray}
	H \equiv \langle \hat{H} \rangle &=& b_2\left( \frac{5n}2 
		+\frac{n^2}2-\frac{3n^2}{2\Omega }\right)+G_0\left( - 
		\frac{n^2}4+\frac{3n^2}{4\Omega }+\frac{n\Omega}
		6\right)  \nonumber \\ 
	& & ~~ +\left( b_2\frac{1-2\Omega }3+G_0\frac{2\left(\Omega -3\right)}
		9\right) H_1+G_0\frac{2\Omega ^2}9H_2 \, ,
\end{eqnarray}
where
\begin{eqnarray}
	H_1&=& \frac{3n}2\left( P_x^2+P_y^2\right) -\Omega \left(
		P_x^4+P_y^4+P_x^2P_y^2\right) \, , \\
	H_2 &=& \sqrt{\left( 1-P_x^2\right) \left( 1-{\cal P}_z^2\right) }
		P_x {\cal P}_z\cos \left( \frac{Q_x}{P_x}\right) 
		\nonumber \\ 
	& & ~~ +\sqrt{\left( 1-P_y^2\right) \left( 1-{\cal P}_z^2\right) }
		P_y {\cal P}_z\cos \left( \frac{ Q_y}{P_y}\right) \nonumber 
		\\ 
	& & ~~ +\sqrt{\left( 1-P_x^2\right) \left( 1-P_y^2\right) }P_xP_y
		\cos \left( \frac{ Q_x}{P_x}-\frac{Q_y}{P_y}\right) \, ,
\end{eqnarray}
with
\begin{equation}
	{\cal P}_z=\sqrt{\left( \frac{3n}{2\Omega }-P_x^2-P_y^2\right) } \, .
\end{equation}
The new Hamilton-like equations become
\begin{eqnarray} 
	\frac{2\Omega }3\dot{P}_i&=&-\frac{\partial H}{\partial Q_i} 
		\nonumber \, , \\ 
	\frac{2\Omega }3\dot{Q}_i&=&\frac{\partial H}{\partial P_i}
	, \qquad i=x,y  \, .  \label{sp6em}
\end{eqnarray}
Note that the 3rd canonical momentum $P_z$ is a conservative quantity
because $Q_z$ is an ignorable coordinate (i.e., it does not appear in the
Hamiltonian function). Therefore, the quantum phase space of nuclear 
deformation is reduced to a four-dimensional space after solving the 
constraint of the fixing valence nucleon number. Thus, we have formulated 
the semiquantal dynamics of the Sp(6) model in a four-dimensional
quantum phase space, which determines the dynamical evolution of nuclear
deformation.

\section{\bf THE DYNAMICAL PAULI EFFECT AND CHAOTIC MOTION}

\subsection{Pauli forbidden region in $( P_i,Q_i)-$phase space}

Although we have formulated the semiquantal dynamics of the Sp(6) model in a
four-dimensional phase space, this phase space is still an nontrivial
geometrical space. As we have seen in the previous section, this quantum
phase space is a compact space. It is embedded in a four-dimensional flat
space. Besides, due to the Pauli effect of the Sp(6) model \cite
{Wu88}, not the whole domain of this quantum phase space is dynamically
allowed. When the valence nucleon number $n \geq {\frac{2\Omega }{3}}$, 
some region is forbidden due to the Pauli effect \cite{zhang89}. 
Explicitly, we can see from Appendix A that the diagonal submatrix block 
parameters ${\bf Y}$ in the Sp(6) CS satisfy the condition ${\bf 
Y}^2={\bf 1}-{\bf XX}^{+}$. Consider the parameter space in the 
intrinsic frame. The matrix ${\bf Y}_{in} ( =\pm \sqrt{{\bf 1}-{\bf 
X}_{in}{\bf X}_{in}^{+}} ) $ has the form: 
\begin{equation}
	{\bf Y}_{in}{\bf =\pm }\left( \begin{array}{ccc}
		\sqrt{1-\left| S_x\right| ^2} & 0 & 0 \\ 
		0 & \sqrt{1-\left| S_z\right| ^2} & 0 \\ 
		0 & 0 & \sqrt{1- \left| S_y\right| ^2} \end{array} \right) \, .
\end{equation}
Since the matrix ${\bf Y}_{in}$ is hermitian, it means that $|S_i|^2
\leq 1$. Together with the condition of the fixing valence
nucleon number, $\sum_i |S_i|^2=\frac{3n}{2\Omega }$, we 
obtain the following conclusion: When $\Omega \geq n > \frac{2\Omega 
}3$\ $\left( \text{or } \sum_k | S_k|^2 > 1\right) $ 
[see Eq.~(\ref{fvnm})], the condition $| S_i| \leq 1$ is not satisfied 
for the whole domain. A forbidden region appears in the quantum phase 
space, and its contour plotted in $\alpha _{20}-\sqrt{2} \alpha _{22}$ 
plane can be determined by using Eq.~(\ref{B2}) \cite{zhang89}. When 
$n \leq \frac{2\Omega }3$\ $\left( \text{or }\sum_k | S_k|^2 \leq
1\right) $, the condition $| S_i| \leq 1$ is always satisfied so that
there is no forbidden region in the phase space. Note that the critical 
value $n=\frac{2\Omega }3$ has an important consequence in the Sp(6) 
model.  When $n \leq {\frac{2\Omega }{3}}$, the nuclear collective motions 
can be described in the symmetric representation of the subgroup SU(3) of 
Sp(6), which behaves the same as the bosonic SU(3) representation. However, 
when $\Omega \geq n > {\frac{2\Omega }{3}}$, the Pauli principle forces 
the system to move to an asymmetric representation of SU(3). Correspondingly, 
the nuclear deformation will be limited in the quantum phase space.  
The condition $| S_i| \leq 1$ is indeed such a phase space 
constraint of the Pauli effect in the Sp(6) model \cite{zhang89}.

In conclusion, within the constraint of the fixing valence nucleon number 
and the presence of the Pauli effect, the quantum phase space of the Sp(6)
model is only a limited domain in a four-dimensional flat space restricted 
by $\frac{3n}{2\Omega }-P_x^2-P_y^2\geq 0$ and $1-|S_i|^2\geq 0$. 
This is a pure quantum mechanical effect on 
geometry. In the following, we shall explore the dynamical consequence 
of these quantum effects to the chaotic motions.

\subsection{Analysis of chaotic motion in the Sp(6) model}

In this subsection, we shall numerically analyze the chaotic motion 
in the Sp(6) model. Since the semiquantal dynamics is determined by the
Hamilton-like equations of motion (\ref{sp6em}), the chaos can be 
described in terms of Poincare' surface of section (PSS). We will first
examine the nuclear surface motions in the dynamical symmetry limits 
of the Sp(6) model. The Hamiltonian in the intrinsic frame can be 
expressed as: 
\begin{eqnarray}
 	H &=& G_0 \langle \hat{S}^{+} \hat{S}\rangle 
		+b_2\sum_{l=1,2} \langle \hat{P}_l\cdot \hat{P}_l\rangle 
		\nonumber \\ 
	& \equiv& T_{eff}(\alpha _{\lambda \mu },\pi _{\lambda \mu })+V_{eff}
	(\alpha_{\lambda \mu }) \nonumber \\ 
	& =& T_{eff}(P_i,Q_i)+V_{eff}(P_i) \,,
\end{eqnarray}
where $T_{eff}$ is an effective kinetic energy which contains the general
kinetic energy and a velocity-dependent potential; and $V_{eff}$ is an
effective potential of the Sp(6) model derived in Appendix B. In the 
intrinsic frame, the Sp(6) model has two group chains: the SU(2) chain
(with $b_2=0$) and the SU(3) chain (with $G_0=0$). These two limiting 
cases describe the nuclear vibrational mode and the $\gamma $-stable 
rotational mode, respectively. The 
numerical results of Fig. 1a and 1b shows that their PSS are regular. 
This is because the existence of dynamical symmetries in these two limits
guarantees the integrability of dynamics \cite{zhang88a}. However, when
both $G_0$ and $b_2$ are not equal to zero, the above dynamical symmetries
are broken and chaos appears. This is shown in Fig. 1c.

To demonstrate the chaotic behavior in the nuclear surface motions, we 
fix the rate $G_0/b_2$ and vary the particle number $n$ to analyze the 
phase space structure of the semiquantal dynamics. There is a phase 
transition from the vibrational to the rotational modes in the nuclear 
surface motions. According to the result in Ref. \cite{zhang88}, this 
phase transition is associated with the dynamical symmetry breaking from 
the SU(2) to the SU(3) limit with fixed $G_0/b_2$, and taking 
$n$ as the control 
parameter. Therefore, chaos must occur accompanied with this 
phase transition \cite{zhang88a,zhang91}. In the following numerical 
calculation, we take $G_0=-.035$ MeV and $b_2=-.065$ MeV with $\Omega = 
36$. Note that the chaotic behavior also varies with the energy 
which can be regarded as a scale for analyzing the strength of
chaotic motion.

The numerical results show that the patterns of PSS in the Sp(6) model
can be divided into three types with $n$ increasing. Foe the cases 
of $n=2\sim 8$, the PSS have the `similar' patterns with the energy 
increasing. We classify this type of PSS class A. The cases $n=10
\sim 16$ form class B, and $n=18\sim 36$ form class C. Here, 
the word `similar' means that the PSS are topologically similar, namely,
the PSS of two states have the similar type of KAM surfaces and the 
similar distribution of chaotic phase space structure. For example, in 
Fig. 2c and 2d, the state of $E=-6.8$ MeV with $n=10$ is similar to the 
state of $E=-7.7$ MeV with $n=12$. Analogously, the state $E=-3.95$ MeV 
with $n=10$ is similar to $E=-4$ MeV with $n=12$, and $E=-3.4$ MeV with 
$n=10$ is similar to $E=-3.5$ MeV with $n=12$. Physically, this 
classification depends on the contribution rate of the quadrupole
and the pairing interactions. Roughly speaking, class A corresponds 
to the states with pairing interaction dominating, class C with 
quadrupole interaction dominating, and class B is the mixed 
states. In Fig. 2, six special cases of the PSS are plotted. They 
show the differences of PSS among the classes A, B and C. These Poincare 
surfaces of section (with $Q_x=0)$ are plotted in the $\alpha
_{20}-\alpha _{22}^{^{\prime }}$ plane $\left( \alpha _{22}^{^{\prime
}}\equiv \sqrt{2}\alpha _{22}\right) $, and have a $\gamma =\frac \pi 3$
-axis symmetry. 

In order to define chaotic strength in a uniform way, we rescale the time 
$t\rightarrow t V_{max}$ [$V_{max}= ( \max \text{imum of }| V_{eff}|)$] 
such that the minimum values of the allowed energy are rescaled to $-1$ 
for all the values of $n$. This rescaling does not change the structure of 
the corresponding PSS. Also, for convenience, we introduce three energy 
scales: $E_u$, $E_w$ and $E_s$. The energy scale $E_s$ is the starting 
energy of the strong quantum chaos, $E_w$ the starting energy of the 
weak quantum chaos, and $E_u$ the upper bound of energy within which 
the Sp(6) model dominates the low-lying nuclear dynamics. Thus, 
for a given nucleus, 
the dynamical behavior of physical states can be divided into three types 
of motions: The pure regular motion which exists in the energy range from 
$-1$ to $E_w$, the weak chaotic motion in the range of $E_w$ to $E_s$, 
and the strong chaotic motion in the range of $E_s$ to $E_u$. Recalling 
the `pattern classification' of the chaotic behavior aforementioned, 
systems with different $n$ in the same class have similar patterns
of PSS. Thus, $E_w$, $E_s$ and $E_u$ are all $n$ dependent. We plot 
the phase diagram of chaotic behavior in the nuclear surface motions 
in the $E$--$n$ plane in Fig. 3. The chaotic strength is obtained
by comparing the values of $E_s$ and $E_w$ in different $n$.

From Fig. 3, we see that when $n$=2 and 4 (small $n$), the contribution
of the quadrupole interaction $( b_2\sum_{l=1,2}\langle \hat{P}
^l\cdot \hat{P}^l\rangle \text{ in Eq.~(\ref{eq33})}) $ is 
much smaller than the pairing interaction $( G_0 \langle\hat{S}^{+}
\hat{S} \rangle) $. This implies that the system is near in the SU(2)
$\subset$ Sp(6) dynamical symmetry. Correspondingly, the nuclear surface
motion is near-integrable. Hence, for small $n$, the nuclear deformation 
is regular in all the energy ranges, as shown in Fig. 3 where $E_w
\simeq E_s\simeq E_u$ for small $n$. This corresponds to class A. 
When $n$ is near $\frac{2\Omega }3$, we find that the quadrupole 
interactions is dominant. The nuclei are approximately described by 
the SU(3) $\supset $ Sp(6) dynamical symmetry. As we can see from
Fig. 3, $E_w$ and $E_s$ are far from the minimum energy $-1$. Therefore 
the phase space is filled mostly by the regular motion as well (with 
$n=18\sim 24$). This corresponds to class C. When the particle number 
$n$ is between 8 to 14, the pairing interaction and the quadrupole 
interactions compete against each other, and all the dynamical 
symmetries in the Sp(6) model are broken. This is the transition 
(mixed) region from the vibration to rotation in nuclear collective 
motions. The chaotic motion is very strong as it is shown in Fig. 3. 
This corresponds to class B.

The most remarkable result we find in this study is in the region 
${\frac{2\Omega }3}<n<\Omega $. Naively, when $n>{\frac{2\Omega }3}$,
the quadrupole interaction should be dominant so that the system should
approximately have the SU(3) dynamical symmetry. However, the numerically 
result shown in Fig. 3 tells us that when $n$ varies from $\frac{2\Omega 
}3$ to the middle shell (${\frac \Omega 2}$), the strength of chaotic 
dynamics is increased. This unusual result is indeed caused by the 
Pauli effect. The Pauli principle forces the system to jump from the 
symmetric SU(3) representation to the antisymmetric SU(3) representation 
when $n>{\frac{2\Omega }3}$. This reduces the nuclear deformation 
(inhibiting the quadrupole interaction and increasing pairing 
interaction). As a result, the system deviates away from the SU(3) 
dynamical symmetry and therefore leads to the stronger chaos. This
is the first example to dynamically examine the Pauli effect on chaos.
It is also the first evidence of showing the enhancement of chaos in 
quantum system due to the Pauli principle.

\section{\bf SUMMARY}

In this paper, the Sp(6) model have been chosen as an example to examine
the chaotic dynamics in quantum systems caused by dynamical symmetry
breaking and dynamical Pauli effect. It has been tested for many systems
where dynamical symmetry breaking is closely related to quantum chaos 
\cite{zhang95,zhang88a,zhang91}. However, this is the first paper to 
address the problem of Pauli effect on chaotic dynamics. Since Pauli
effect is a pure quantum mechanical phenomenon and it has no classical 
limit, its influence on chaos is certainly an important key to 
understand the intrinsic mechanism of quantum chaos. In this 
paper, we only present the evidence of Pauli effect on chaos in a 
specific system. The most interesting and challenge questions are 
whether the dynamical Pauli effect can always enhance chaotic motions
and what generic role does the Pauli principle play in nonlinear 
dynamics of quantum theory? We leave these questions for future 
investigations.

\begin{center}
{\bf ACKNOWLEDGMENTS}
\end{center}

We would like to acknowledge the discussions with D. H. Feng, H. T. Mo, Y.
H. Wong, S. L. Lin and C. T. Li. We also thank S. P. Li for his careful 
reading of the manuscript. This work is supported by the Department
of Physics of National Taiwan University under Grant 5010202-02-434 
(MTL) and the National Science Council of the Republic of China under 
Grant NSC86-2816-M001-008L (WMZ).

\newpage

\appendix

\section{Sp(6) Coherent States}

To make this paper self-contained, we briefly list in this appendix
the main results of the Sp(6) coherent states which has been discussed 
in Ref. \cite{zhang88}. The coherent states of the group USp(6) (the 
compact group of the Lie algebra Sp(6)) are defined by
\begin{equation}
	| \eta \rangle =\hat{T}{\bf (}\eta {\bf )} | 0 \rangle 
		\, ,  \label{1}
\end{equation}
where the generalized displacement operator $\hat{T}{\bf (}\eta {\bf )}$,
the coset representative of USp(6)/U(3), is given by

\begin{equation}
	\hat{T}( \eta )=\exp \left[ \left( \eta _{00}\hat{S}{\bf 
		^{+}+}\sum_\mu \eta _{2\mu }\hat{D}_\mu ^{+}\right) 
		-H.C.\right] \, ,  \label{A2} 
\end{equation}
and $| 0 \rangle$ is the core vacuum (nearest closed shell), $\hat{S}{\bf 
^{+}}$ and $\hat{D}_\mu ^{+}$ the pair creation operators in the Sp(6)
model. The matrix representation of $\hat{T}(\eta)$ has the form:
\begin{equation}
	\hat{T}(\eta) \rightarrow \text{Rep} (\hat{T}) = \left[ 
		\begin{array}{cc} {\bf Y} & {\bf X} \\ 
			-{\bf X}^{+} & {\bf \tilde{Y}} \end{array}
		\right] \equiv {\bf T} \, ,  \label{A3}
\end{equation}
where the matrices ${\bf X}$ and ${\bf Y}$ are the 3$\times $3 subblock 
matrices of the coset representative of USp(6)/U(3). Their explicit forms 
are given by
\begin{equation}
	{\bf X}=\left[  \begin{array}{ccc}
		\sqrt{3}\chi _{22} & \sqrt{\frac 32}\chi _{21} 
			& \chi _{00}+\sqrt{\frac 12} \chi _{20} \\ 
		\sqrt{\frac 32}\chi _{21} & -\chi _{00}+\sqrt{2}
			\chi _{20} & \sqrt{\frac 32} \chi _{2-1} \\ 
		\chi _{00}+\sqrt{\frac 12}\chi _{20} & \sqrt{\frac 32}
			\chi _{2-1} & \sqrt{3} \chi _{2-2} \end{array}
		\right] \, ,  \label{A4}
\end{equation}
where $\chi_{\lambda \mu }$ is a function of $\{\eta_{\lambda \mu} \}$.
There are totally twelve parameters in this coherent state. Since the 
matrix {\bf T} is unitary, the matrix {\bf X} must be symmetric and 
the matrix {\bf Y} is hermitian and is determined by ${\bf X}$ from 
the following relations:
\begin{equation}
	{\bf YY+XX^{+}=I} ~~, ~~~ {\bf YX=X\tilde{Y}} \, . \label{A5}
\end{equation}

\section{Geometrical model via the Sp(6) model}

In this appendix, the geometrical model and the Sp(6) model will be 
compared. Without any loss of generality, we define the nuclear surface 
variables as
\begin{eqnarray}
	\alpha _{20}(t)&=& \beta (t)\cos [\gamma (t)]\quad \nonumber \, ,\\ 	
	\alpha _{22}(t)&=&\frac 1{\sqrt{2}}\beta (t)\sin [\gamma (t)]
		\, , \label{B1}
\end{eqnarray}
where $\beta (t)$ and $\gamma (t)$ are the usual geometrical deformation
parameters. Substituting Eq.~(\ref{B1}) into Eq.~(\ref{eq23}), it gives 
\begin{equation}
	| S_i| = -\alpha _{00}+\sqrt{2}\beta \cos \left( \gamma 
		-\frac{2k_i }3\pi \right) \, ,\label{B2}
\end{equation}
where $i=z,x,y; k_z=0, k_x=1, k_y=2$ and
\begin{equation}
	-\alpha _{00}=\sqrt{\frac n{2\Omega }-\beta ^2}  \, . \label{B3}
\end{equation}

The effective potential of the Sp(6) model is 
\begin{equation}
	V_{eff}=\langle \hat{H}(\pi _{00},\pi{20},\pi _{22}=0;$ $\alpha _{00},
		\alpha _{20},\alpha _{22}) \rangle \, ,
\end{equation}
where $\pi_{\lambda \mu }$ is the conjugate momentum of $\alpha 
_{\lambda \mu }.$ When these variables are zero, the variables 
$S_i$ in Eq.~(\ref{eq23}) become real (for a proof, see Appendix C). 
Substituting the real variables $S_i$ of Eq.~(\ref{B2}) into 
Eqs.~(\ref{cseve}-\ref{csev1}), the effective potential can be expressed as:
\begin{equation}
	V_{eff [\text{G}]}=A\beta ^4+B\beta ^2+C\alpha _{00}\beta ^3
		\cos 3\gamma +D+F  \, . \label{B4}
\end{equation}
where the stiffness coefficients $A,B,C,D$ and $F$ are listed as follows:
For G = SU(2),
\begin{eqnarray}
	A&=& 7\Omega (\Omega /3-1)/2\quad ,\quad B=-2n(\Omega /3-1)\quad 
		\nonumber \, , \\ 
	C&=& 2\sqrt{2}\Omega (\Omega /3-1)\quad ,\quad D=n(\Omega -n/2
		+3n/(2\Omega ))/6 \nonumber \, , \\ 
	F&=& (\Omega /3)^2\sum_{i\neq j, i,j=x,y,z} S_iS_j\left[ \left( 
		1-S_i^2\right) \left( 1-S_j^2\right) \right] ^{1/2}
		\, , \label{B5}
\end{eqnarray}
and for G = SU(3),
\begin{eqnarray}
	A&=& -7\Omega (2\Omega -1)/4\quad ,\quad B=n(2\Omega -1) \nonumber 
		\, , \\ 
	C&=& -\sqrt{2}\Omega (2\Omega -1)\quad ,\quad D=5n(2\Omega -n)/
		(4\Omega )\quad ,\quad F=0 \, .\label{B6}
\end{eqnarray}
These results  agree with that of Ref.\cite{zhang88}. The effective 
potential of the Sp(6) model also has the same form as the potential 
energy of the geometrical model in the intrinsic frame
\begin{equation}
	V_{eff}[\beta ,\gamma ]=\frac 15C_4\beta ^4+\frac 12C_2\beta^2
		-\sqrt{\frac 2{35}}\beta ^3\cos 3\gamma +\cdot \cdot 
		\cdot  \, . \label{B7}
\end{equation}
In other words, the Sp(6) model in this picture contains all the properties 
of the nuclear geometrical model.

\section{Deformation phase space}

In this appendix, we will show that the surface variables, $\alpha 
_{00},\alpha _{20}$ and $\alpha _{22}^{^{\prime}} $ and their conjugate 
momentum $\pi _{00},\pi _{20}$ and $\pi_{22}^{^{\prime }}$ form the 
canonical variables, where $\alpha_{22}^{^{\prime }}\equiv \sqrt{2}
\alpha _{22}$. To do so, we introduce the fourth type generating
function 
\begin{equation}
F^{(4)}=-\frac{\pi _{00}}3\left( {\cal P}_x+{\cal P}_y+{\cal P}_z\right) - 
\frac{\pi _{20}}{3\sqrt{2}}\left( {\cal P}_x+{\cal P}_y-2{\cal P}_z\right) - 
\frac{\pi _{22}^{^{\prime }}}{\sqrt{6}}\left( {\cal P}_y-{\cal P}_x\right)
\, .  \label{C2}
\end{equation}
It transforms the canonical variables $\left\{ {\cal P}_i,{\cal Q}_i,
i=x,y,z\right\} $ mentioned in Sec. III into another set of the variables 
$\{\alpha _{\lambda \mu },\pi _{\lambda \mu } \} $. The connection between
these two sets of variables are 
\begin{eqnarray}
	{\cal Q}_x&=&-\frac{\partial F^{(4)}}{\partial {\cal P}_x}
		=\frac 13\left( \pi_{00}+ \frac{\pi _{20}}{\sqrt{2}}
		-\sqrt{\frac 32}\pi _{22}^{^{\prime }}\right) \nonumber , \\ 
	{\cal Q}_y&=&-\frac{\partial F^{(4)}}{\partial {\cal P}_y}
		=\frac 13\left( \pi_{00}+ \frac{\pi _{20}}{\sqrt{2}}
		+\sqrt{\frac 32}\pi _{22}^{^{\prime }}\right) \nonumber , \\ 
	{\cal Q}_z&=&-\frac{\partial F^{(4)}}{\partial {\cal P}_z}=
		\frac 13\left( \pi_{00}- \sqrt{2}\pi _{20}\right) 
		\, ,  \label{C3}
\end{eqnarray}
and $\alpha _{\lambda \mu }=\frac{\partial F^{(4)}}{\partial \pi _{\lambda
\mu }}$ gives the same relation listed in Eq.~(\ref{eq23}). The explicit 
connection between the variables $S_{i\text{ }}$ and ($\alpha _{\lambda \mu 
},\pi _{\lambda \mu }$) can be found by using $S_i={\cal P}_i\exp 
(-i\frac{{\cal Q}_i}{{\cal P}_i})$. Hence $\left\{ \alpha _{\lambda \mu
},\pi _{\lambda \mu }\right\} $ are the canonical variables. Eq.~(\ref{C3})
and $S_i={\cal P}_i\exp (-i\frac{{\cal Q}_i}{{\cal P}_i})$ imply that 
$S_i$ is real when $\pi _{\lambda \mu }=0 $. Thus we can easily obtain
the effective potential discussed in Appendix B.

%\reference

\begin{center}
{\bf FIGURES}
\end{center}

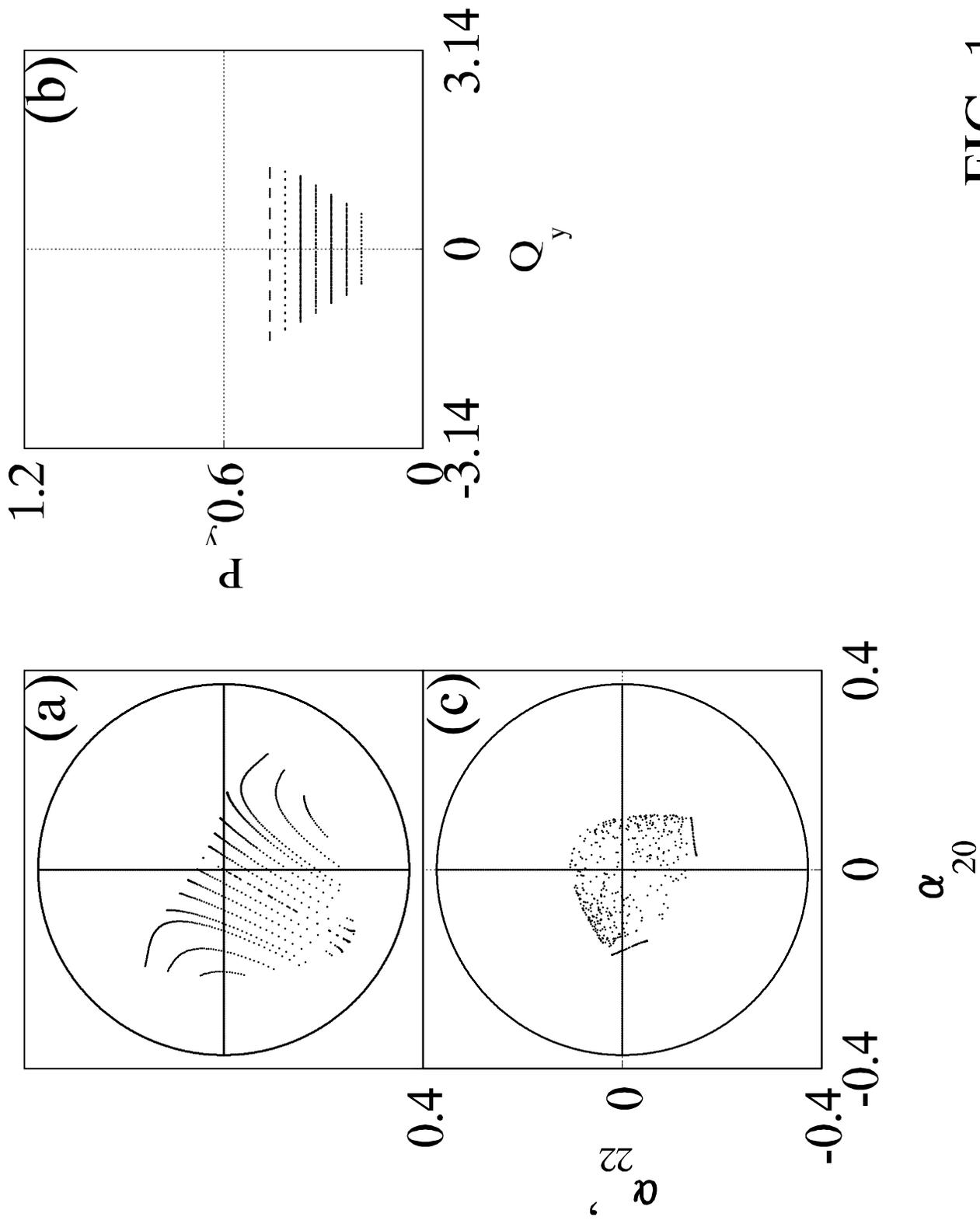
FIG. 1. The Poincare surface of section ($Q_x=0$) of the Sp(6) model with $%
\Omega =36$, $n=10$ and $E=-3.0$ MeV: (a) $G_0=-0.1$ MeV, $b_2=0.$ MeV, (b) $%
G_0=0.$ MeV, $b_2=-.1$ MeV, (c) $G_0=-0.01$ MeV, $b_2=-.09$ MeV.

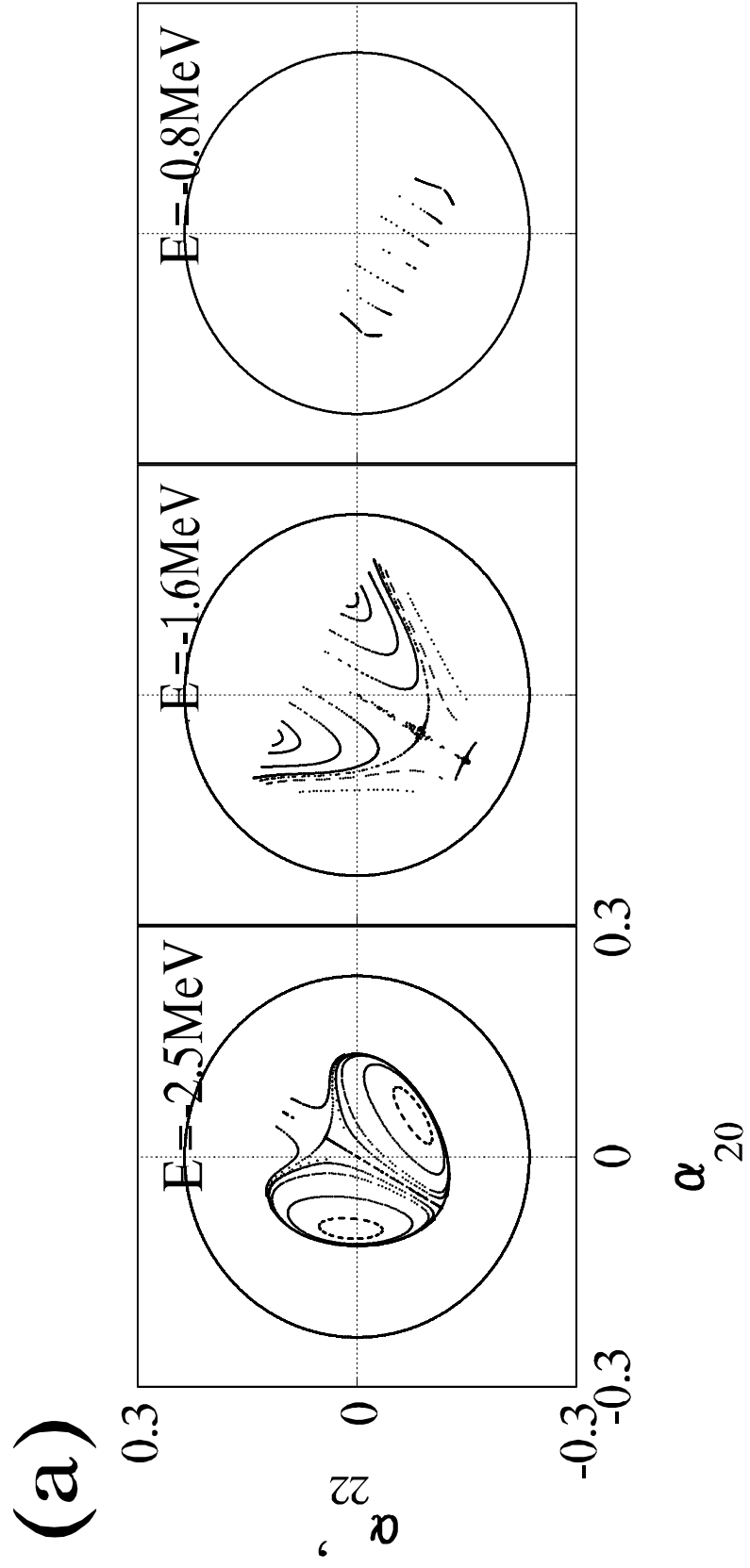
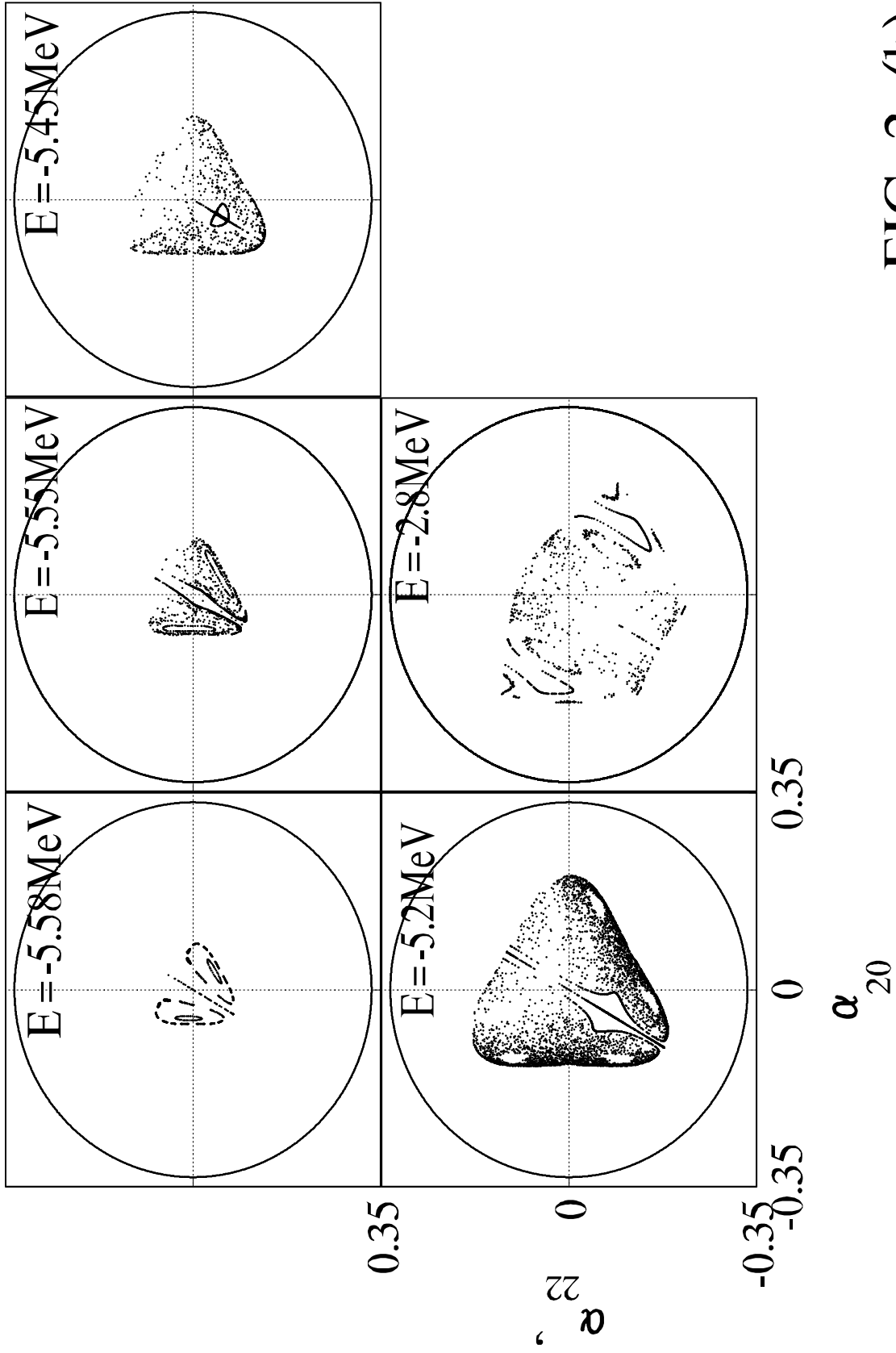
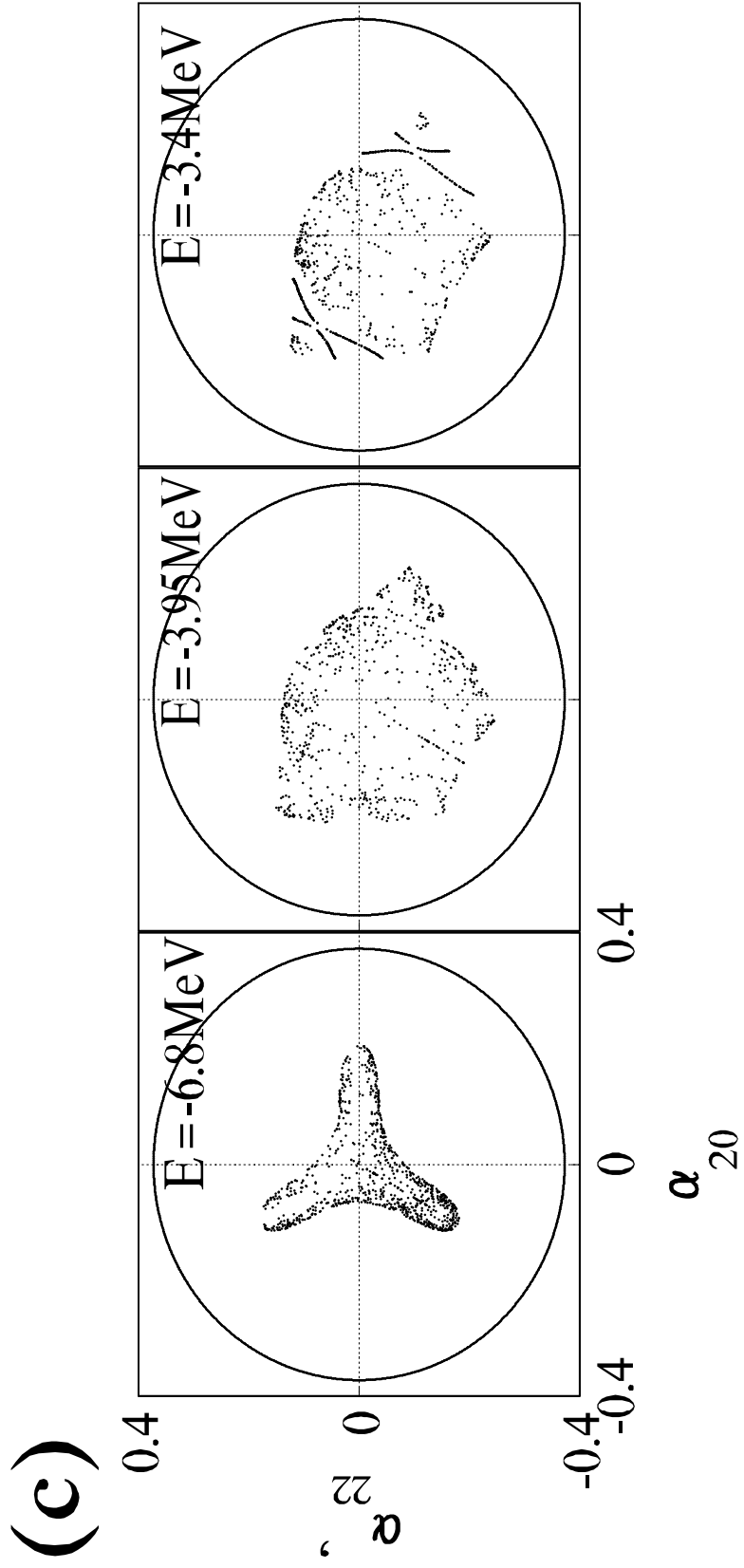
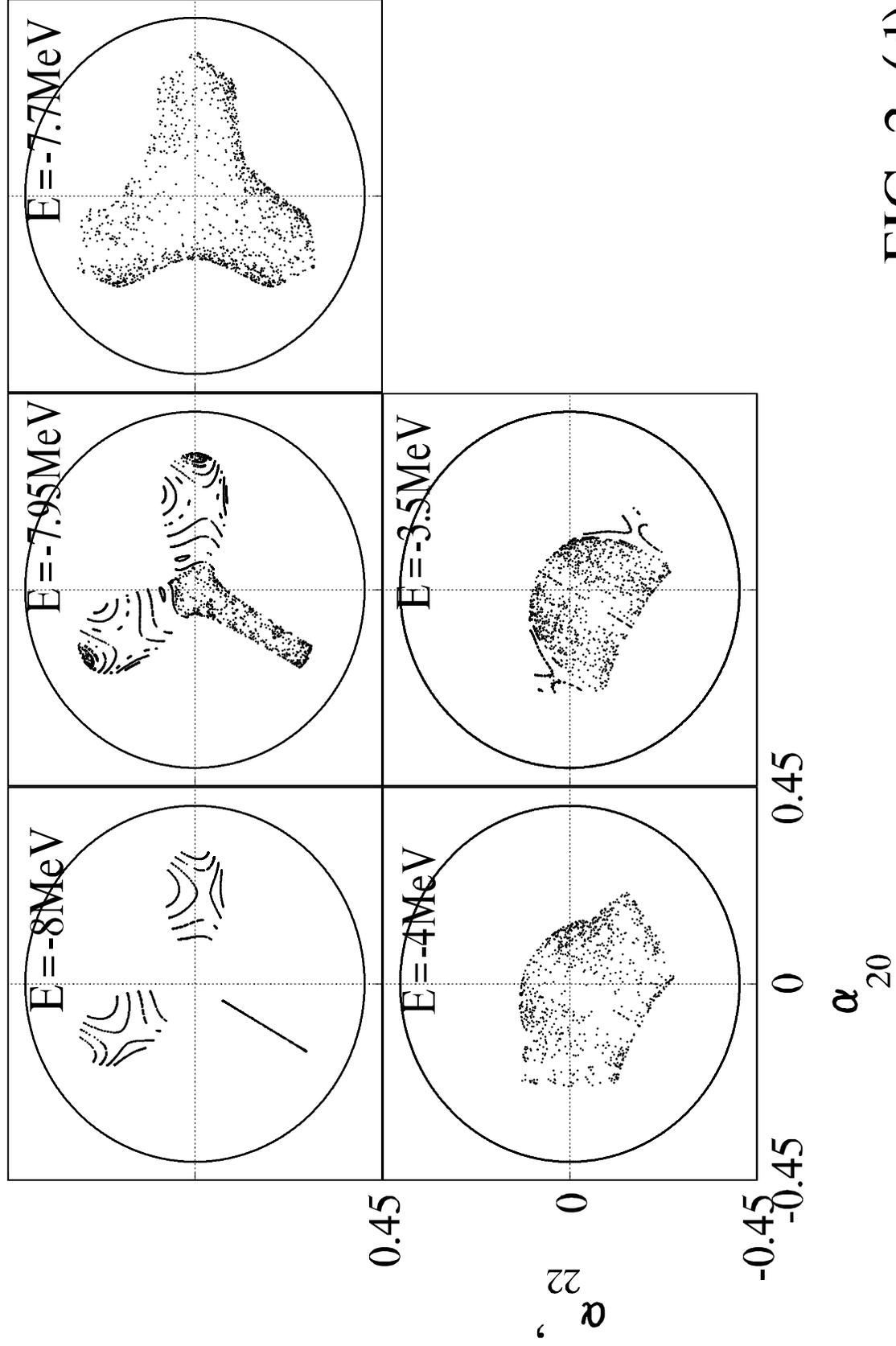
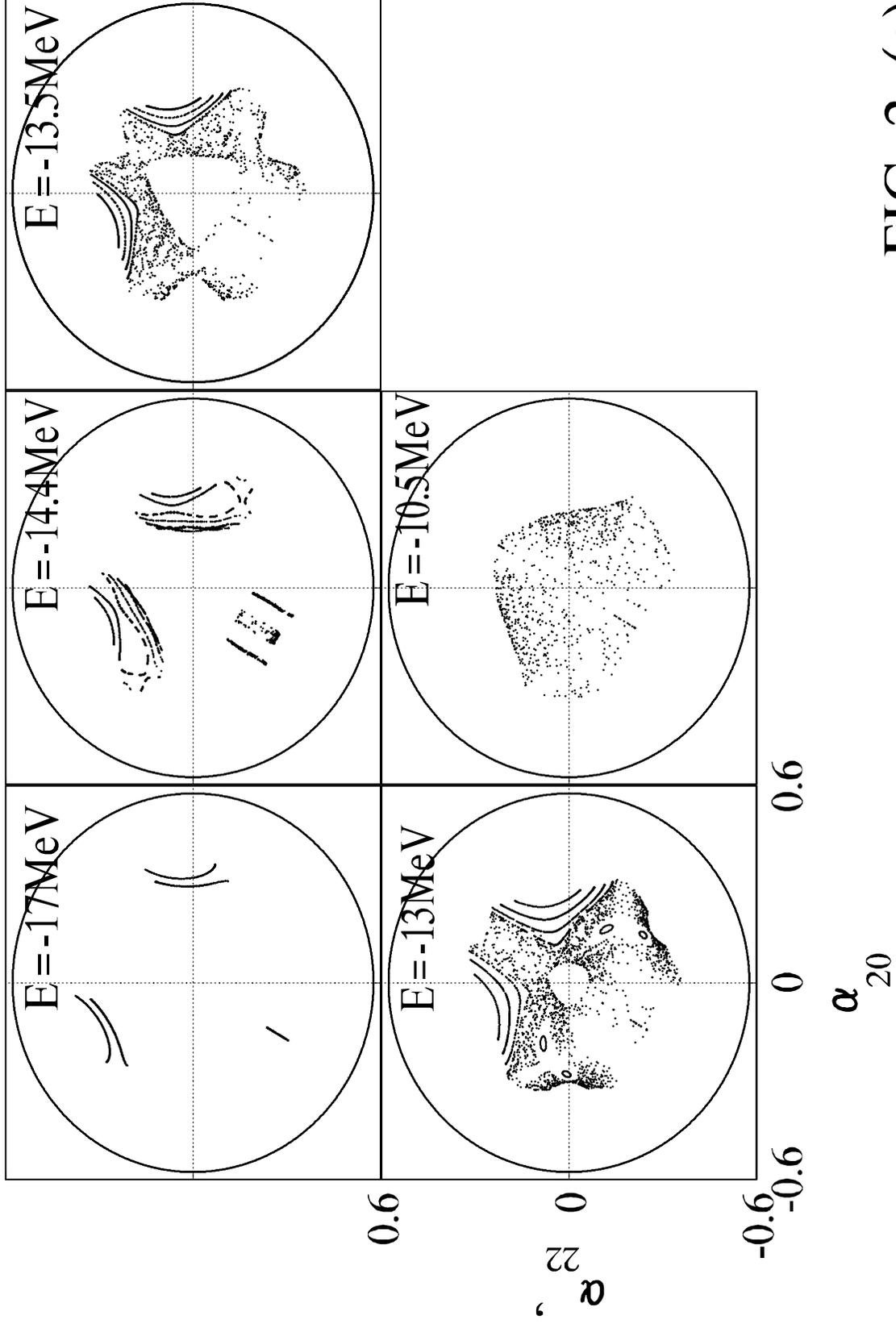
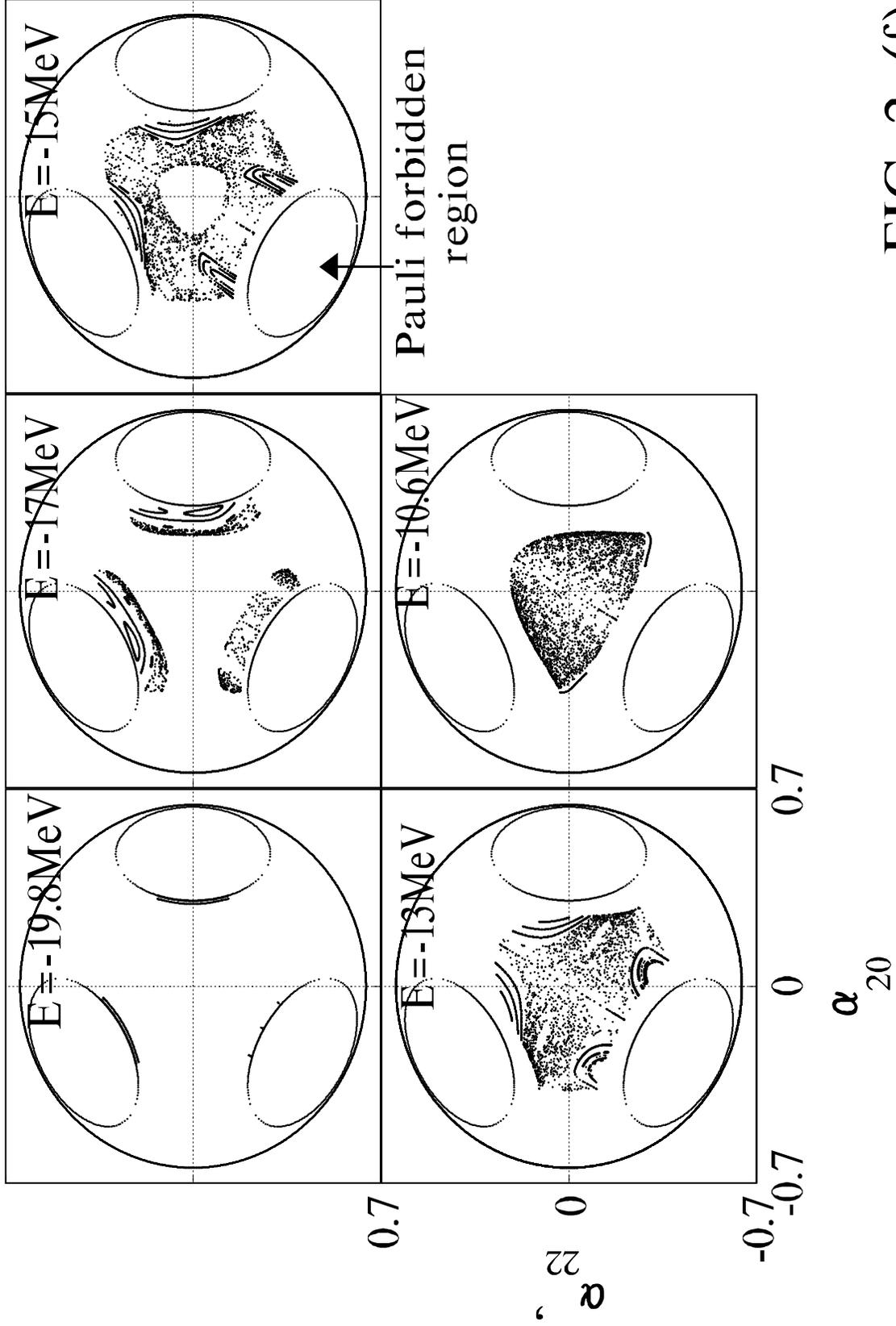
FIG. 2. The Poincare surface of section ($Q_x=0$) of the Sp(6) model with
energy increasing, where $\Omega =36$, $G_0=-.035$ MeV, $b_2=-.065$ MeV: (a) 
$n=4$, (b) $n=8$, (c) $n=10$, (d) $n=12$, (e) $n=24$, (f) $n=30$.

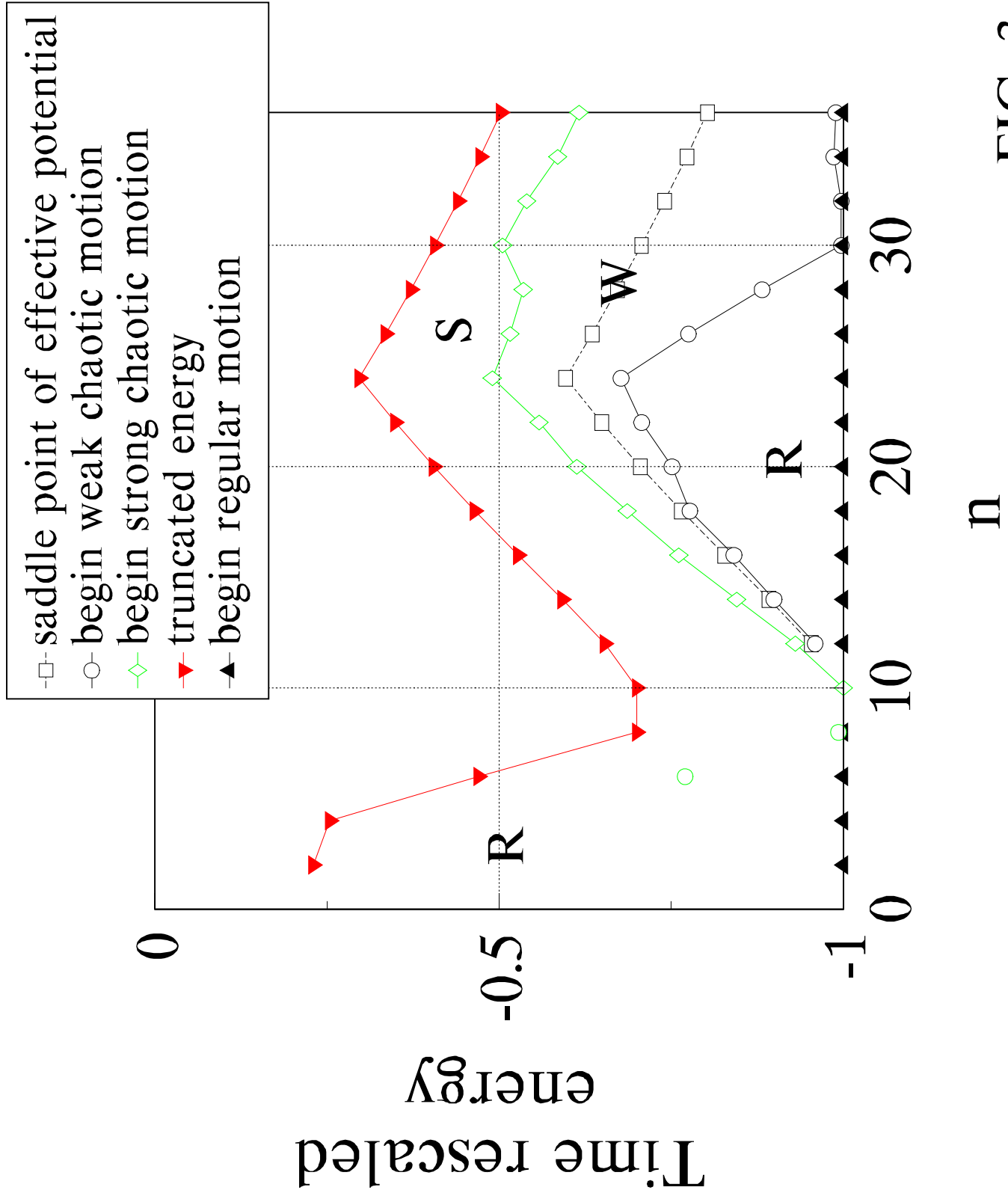
FIG. 3. A phase diagram of chaotic patterns in $E-n$ plane in the Sp(6)
model with $\Omega =36$, where $R$ means the regular region, $W$
the weak-chaotic region, and $S$ the strong-chaotic region.

\end{document}